\documentclass[twocolumn,showpacs,preprintnumbers,amsmath,amssymb]{revtex4}

\usepackage{graphicx}
\usepackage{dcolumn}
\usepackage{bm}

\begin{document}

\preprint{APS/123-QED}

\title{New method of studying slow strange meson
properties in nuclear matter}

\author{Yu.T.Kiselev}
\email{yurikis@vitep1.itep.ru}
\author{V.A.Sheinkman}%

\affiliation{ Institute of
 Theoretical and Experimental Physics,
 Moscow, 117218, Russia }

\date{\today}

\begin{abstract}
We suggest the new experimental method to explore the properties
of slow strange mesons at normal nuclear matter density. We show
that the $K^{+}$ and $K^{-}$ mesons with extremely small momenta
relative to the surrounding medium rest frame can be produced  in
nucleus-nucleon collisions and their production cross sections are
experimentally measurable. The experiments on study of the
momentum dependence of meson-nuclear potentials are discussed.
\end{abstract}


\maketitle The question about the properties of mesons in baryon
matter has attracted much attention during last years \cite{Ref1}.
In particular, the investigation of strange mesons is of special
interest as it is related to a partial restoration of chiral
symmetry \cite{Ref2} and possible existence of an antikaon
condensed phase in dense interior of neutron stars \cite{Ref3}.
According to theoretical studies based on various approaches such
as effective chiral Lagrangians \cite{Ref4} and mean-field models
\cite{Ref5} the in-medium antikaon mass should be substantially
reduced while the kaon mass is expected to be slightly enhanced.
This has triggering a considerable interest in the studying of
$K^-$ meson production in heavy-ion collisions. The dropping of
the $K^-$ mass in nuclear matter has strong impact on the antikaon
yield especially in the subthreshold reactions due to in-medium
shift of the elementary production threshold to lower energies. An
enhanced antikaon to kaon ratio in $Ni+Ni$ collisions in
subthreshold energy regime has been observed at GSI
\cite{Ref6,Ref7} and was attributed to the attractive antikaon
potential of $-(100-120)$ MeV at nuclear saturation density
\cite{Ref8}. A strong attractive optical potential  of about
$-(180-200)$ MeV at normal nuclear density for the $K^-$ has been
extracted  from experimental data on kaonic atoms \cite{Ref9}. The
substantial difference between the antikaon potentials mentioned above,
i.e. their change in the mass, can be understood if the
potential is momentum dependent. Indeed, the experiments with
kaonic atoms deal with the $K^-$ mesons of zero momentum relative
to the nuclear matter rest frame while heavy-ion collisions at GSI
probe the antikaons at momenta of more than 300 MeV/c with respect
to the baryonic fireball \cite{Ref6}. The results of the
calculations performed in the frame of different models
\cite{Ref10,Ref11,Ref12,Ref13,Ref14} show that the influence of an
antikaon potential on the subthreshold production of $K^-$ mesons
increases with lowering of their momenta relative to the
surrounding medium. It is thus desirable to obtain the information
about the $K^-$  and $K^+$ potentials at momentum less than 300
MeV/c. In-medium modification effects are expected to be the most
pronounced in nucleus-nucleus interactions in which the high
density and/or temperature are accessible. However, due to a
complex dynamic involved the interpretation of heavy-ion
experiments is, at present, far from being unambiguous. Therefore,
it is certainly useful to explore such effects under less extreme
but much better controlled conditions at normal nuclear density in
proton-nucleus collisions. However, slow  kaons and especially
antikaons are hard to investigate experimentally in $pA$ reactions
mostly due to their strong decay and small production cross
sections vanishing at zero meson momentum.

We discuss the alternative method of such study. The properties of
mesons with extremely small momenta with respect to the nuclear
matter rest frame can be explored in the inverse kinematics, i.e.
in nucleus-nucleon collisions. As it follows from the Lorentz
transformation a slow particles in a projectile nucleus system
appear to be fast in the laboratory i.e. in the target nucleon
frame of reference. The meson which is at rest inside the incident
nucleus has the same laboratory velocity as a surrounding nucleons. Using of
the inverse kinematics makes possible the investigation of
particle production  in new kinematical range including zero
momentum relative to the nuclear matter rest frame which is not
accessible in $pA$ reactions. In contrast to heavy-ion collisions
the determination of the meson momenta relative to the nuclear
environment in $Ap$ reactions is model  independent.

The suggested method provides the important advantages for an
experimental measurements in the low-momentum range. Large Lorentz
boost with respect to the laboratory results in upward shifts of
the $K^+$ and $K^-$  momenta. As a consequence, their decay losses
significantly decrease and the mesons become convenient for the
detection. Since the invariant cross sections are the same in both
systems an experimentally observed differential cross sections in
the inverse and direct kinematics are related as:

\begin{eqnarray}
(d^{2}\sigma/dpd\Omega)^{Ap}=(p^{2}/E)^{Ap}(E/p^{2})^{pA}(d^{2}\sigma
/dpd\Omega)^{pA}%
\label{eq:one}.
\end{eqnarray}

where $p$ and $E$ stand for kaon (antikaon) momentum and energy
while upper indices denote the type of reaction. For the
production of mesons with low momentum relative to the nuclear
matter rest frame the cross section $(d^{2}\sigma/dpd\Omega)^{Ap}$
considerably exceeds $(d^{2}\sigma/dpd\Omega)^{pA}$ because of the
factor $(E/p^{2})^{pA}$ grows rapidly with lowering of meson
momentum while the factor $(p^{2}/E)^{Ap}$ changes rather
smoothly.

In the present Letter we shall consider some applications of the
suggested method.

\section{Study of the in-medium kaon potential}
\begin{figure}[!h]
\begin{center}
 \includegraphics[width=85mm]{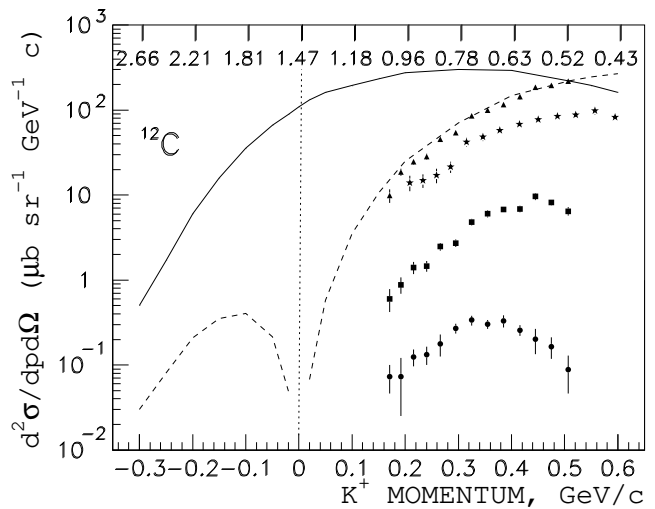}
 \end{center}
 \caption{Double differential $K^+$ production cross sections.
Circles denote the cross sections measured in proton-carbon
collisions at $T_p=1.0$ GeV, squares at $T_p=1.2$ GeV, stars at
$T_p=1.5$ GeV and triangles at $T_p=2.0$ GeV \cite{Ref16}. Dashed
line is our calculation for $T_p=2.0$ GeV in the direct
kinematics. Solid curve represents the corresponding cross section
for $K^+$ production in carbon-proton collisions at the ion beam energy 2 AGeV.} \label{fig:1}
\end{figure}

Let us consider $K^+$ meson production in the inverse kinematics
by an ion beam on hydrogen target. Our estimate of the cross section 
for kaon production in $Ap$
collisions is based on the data from $pA$ reactions as well as the
calculations performed in the framework of simple folding model
\cite{Ref15} disregarding any potentials. Within this approach the
inclusive $K^+$  production in proton-nucleus collisions at near
threshold and subthreshold energies is analyzed with respect to
the one-step $(pN \rightarrow K^+ YN, Y= \Lambda,\Sigma)$  and
two-step $(pN_1 \rightarrow \pi NN, \pi N_2\rightarrow  K^+ Y$)
incoherent processes. The invariant cross sections for both
forward and backward kaon production in $pA$ collisions at initial
proton energies 2 GeV were found taking into account both reaction
channels and then transformed into noninvariant double
differential form. In Fig.~\ref{fig:1} the calculated double
differential cross section for forward $K^+$ meson production on
carbon target (dashed curve) is compared to that measured in the
angular range $0-10^o$ at the same initial proton energy
\cite{Ref16}. Experimental data are seen to be reproduced
reasonably. The simple kinematical calculation
shows that at the beam energy 2 AGeV the kaons produced inside a
projectile nucleus with momenta from zero to 0.3 GeV/c in backward
hemisphere relative to the beam direction appear in the laboratory
within the longitudinal and transverse momentum ranges $0.78\le
P_l\le 1.47$ GeV/c and $0.0\le P_t \le0.3$ GeV/c, respectively.
The $K^+$ mesons emitted in forward hemisphere with momenta up to
0.3 GeV/c will be observed in the laboratory momentum range $1.47
\le P_l \le 2.66$ GeV/c and in the same interval of $P_t$ .
Laboratory momenta (in GeV/c) of the $K^+$ mesons from
carbon-proton collisions corresponding to the momenta of kaons
produced in forward and backward directions in $p^{12}C$ reaction  are plotted on the
upper axis. The solid curve in the figure depicts the double
differential cross section for $K^+$ production in the inverse
kinematics obtained from the calculated cross section (dashed
curve) by using Eq.~\ref{eq:one}. Left-hand part of the figure is
related to an unexplored experimentally range of backward kaon
production in proton-nucleus collisions. The comparison of the
data \cite{Ref16} presented in Fig.~\ref{fig:1} with predicted
cross section shows that it is definitely acceptable for
measurements and significantly exceed that in traditional
kinematics in the most interesting range of low $K^+$ momenta
relative to surrounding nuclear matter. Furthermore, an upward
shift of a kaon momentum in $Ap$ collisions results in sizable
decrease of a $K^+$  decay losses. In contrast to $pA$ reactions
large value of the cross section and its smooth behavior provide
the favorable experimental conditions for the investigation of
in-medium effects in the inverse kinematics. The influence of kaon
nuclear and Coulomb potentials should lead to the deviation of the
cross section from the solid curve calculated without above
potentials. The signature of the effect will be A-dependent dip in
the cross sections at laboratory kaon momentum around 1.47 GeV/c
corresponding to zero $K^+$  meson momentum relative to the
projectile nucleus system.

The evidence for the action of Coulomb and nuclear potentials on
soft kaon production was obtained in \cite{Ref17} where the ratio
of forward $K^+$ meson yield from copper, silver and gold targets
to that on carbon has been measured at proton beam energies
between 1.5 and 2.3 GeV. It was found that in the momentum range
from 170 till 600 MeV/c the ratios exhibit similar shape rising
with decreasing kaon momentum, passing a maximum and falling down
at momenta less than 250 MeV/c. The magnitude of the $K^+$ -
nucleus repulsive potential was found to be 20 MeV at normal
nuclear density. The authors plan to extend the measurements of
the ratios down to kaon momentum of about 100 MeV/c what is not
trivial experimental task. In the inverse kinematics both forward and backward
production ranges in the nucleus rest frame can be
studied simultaneously because of all produced kaons are peaked forward
in the laboratory. As a consequence, the momentum dependence of the ratio of 
$K^+$ production cross section measured with heavy projectile ions
to that measured with light ones should exhibit distinct two peak
structure.

\section{Study of the in-medium antikaon potential}
\begin{figure}[!]
\begin{center}
 \includegraphics[width=65mm]{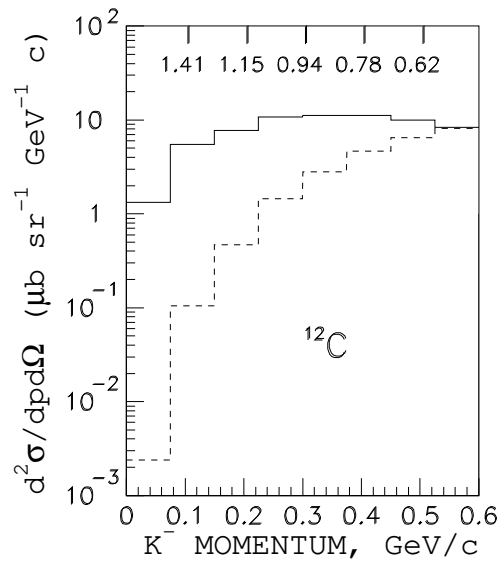}
 \end{center}
\caption{Double differential $K^-$ production cross sections in
the direct (dashed histogram) and inverse (solid histogram)
kinematics.}\label{fig:2}
\end{figure}

Let us consider $K^-$ meson production by an ion beam of energy
2.5 AGeV on hydrogen target. The $K^-$  mesons produced in
backward hemisphere relative to the beam direction with momenta up
to 0.3 GeV/c in the projectile nucleus rest frame will be observed
in the laboratory momentum ranges $0.94 \le P_l \le 1.74$ GeV/c
and $0.0 \le P_t \le 0.3$ GeV/c. This process corresponds to
low-momentum $K^-$ production in forward hemisphere by protons on
nuclear target. The available experimental information about
subthreshold antikaons from proton-nucleus collisions is very
scarce \cite{Ref11,Ref18}. Data on the production of $K^-$ mesons
with small momenta are completely missed now. Under these
circumstances we have to rely upon the model calculations to
evaluate the respective cross section. The forward $K^-$ mesons
production in the $p+A \rightarrow K^- + X$ reaction at the proton
beam energy 2.5 GeV was studied in \cite{Ref10} within a coupled
channel transport approach. The dashed histogram in
Fig.~\ref{fig:2} shows the $K^-$ momentum spectrum for $^{12}C$
target calculated in \cite{Ref10} with zero potentials but taking
into account the antikaon absorption in its way out through the
nucleus. The solid histogram in the figure depicts the double
differential cross section for the $K^-$ meson production in
$^{12}C+p$ collisions obtained from the dashed histogram by using
Eq.~\ref{eq:one}. The upper scale represents the corresponding
laboratory momenta (in GeV/c) of the $K^-$ mesons from
carbon-proton collisions. Taking into account the values of the
antikaon production cross sections and sizable decrease of their
decay losses we conclude that the expected event rate in low $K^-$
momentum range is acceptable for measurements in the $Ap$
kinematics.

Impact of the surrounding medium on slow $K^-$ production should
differ from that on $K^+$  due to an attractive nature of both
Coulomb and nuclear potentials. The action of Coulomb potential
will populate the low-momentum range while the influence of the
nuclear potential depends sensitively on its strength. One can
expect that in the case of weak potential the yield of the $K^-$
mesons will be suppressed due to their strong absorption via the
$K^-+N\rightarrow \Sigma+\pi$ reaction which has very large cross
section at small antikaon momentum. On the contrary, in the case
of strong potential exceeded 100 MeV the $K^-$ mesons
absorption plays a minor role because of the above process is
energetically almost closed \cite{Ref19}. This will lead to an enhanced
low-momentum $K^-$  meson yield even from heavy nuclei. The 
calculations \cite{Ref10} with attractive antikaon potential 120 MeV give
a factor of about 10 enhancement in the cross sections for low
momentum $K^-$ . It is thus
necessary to explore this question experimentally by measuring in
the inverse kinematics the $K^-$ spectra as a function of the
projectile nucleus mass.

If an attractive antikaon-nucleus potential turns out to be large
it will be a strong argument for existing of narrow discrete
nuclear antikaon bound states  (see \cite{Ref19} and references
therein). The using of the inverse kinematics can be  the
promising way to produce such a states.

It should be noticed that slow pions inside nuclei can also be
explored in the inverse kinematics. Such a measurements may help
to disentangle the effects of the Coulomb and nuclear potentials
on kaon and antikaon production.

\section{Study of the subthreshold reaction mechanism}
As it was mentioned above the
kaon and antikaon production in proton-nucleus interactions below the free
NN threshold is reduced to the one-step and two-step processes due to 
the lack of collision energy. It is commonly
believed that the reaction mechanism can be identified from the
target atomic mass dependence of the cross sections. The
A-dependence for the one-step process is determined by the total
inelastic cross section and scales as $\approx A^{2/3}$ provided
weak final state absorption of the produced mesons. The stronger
dependence $\approx A^{1}$ is expected for the two-step kaon
creation process since the respective  cross section includes the
additional probability of the second collision of an intermediate
pion with another target nucleon which is proportional to $\approx
A^{1/3}$. The total $K^+$ production cross section on different
nuclei have been measured in \cite{Ref20} at the proton energy 1 GeV
which is far below the free $NN$ threshold ($T_{NN}=1.58$ GeV).
Note that low-momentum kaons give the main contribution to the
total cross section. By fitting to the cross sections with an
$A^{\alpha}$ function the exponent $\alpha$ was found to be
$1.04\pm0.03$. The strong A-dependence has been interpreted in
\cite{Ref20,Ref21} as an evidence for the dominance of the
two-step reaction mechanism. Recently ANKE Collaboration obtained
the data on double differential cross sections for low-momentum
$K^+$ production on nuclear targets from carbon to gold at the
same proton energy 1 GeV \cite{Ref22}. The extracted value of
$\alpha=0.74\pm0.05$ is in reasonable agreement with A-dependence
expected for the one-step mechanism. This discrepancy does not
allow one to draw unambiguous conclusion about the underlying
reaction mechanism of slow $K^+$ production in $pA$ collisions \cite{Ref23}.

Investigation of low momentum kaon in the inverse kinematics seems
the most promising way to clarify the situation. Using of
different ion beams provides the possibility to explore the atomic
mass dependence. At projectile energy 1 AGeV the $K^+$  meson
emission in backward  hemisphere relative to the ion beam
direction within the momentum range $0.0 - 0.3$ GeV/c in the
projectile nucleus rest frame looks like forward kaon production
in the laboratory momentum intervals $0.43 \le P_l \le 0.89$ GeV/c
and $0.0 \le P_t \le 0.3$ GeV/c. This makes the measurements in
the inverse kinematics more favorable than those in the direct
kinematics. For instance, since the $K^+$ meson momentum 0.1 GeV/c
relative to projectile nucleus corresponds to the laboratory
momentum 0.7 GeV/c, the differential cross section in $Ap$
collisions is enhanced by a factor of 30 (Eq.~\ref{eq:one}).
Furthemore, at the distance between the production target and the
detectors of about 2.5 meters corresponding to the actual
experimental situation, the loss of the kaons due to their decay
in flight is decreased by more than an order of magnitude.
Therefore, the $K^+$ event rate would exceed that in traditional
kinematics by about a factor of  400.
The corresponding enhancement in the event rate of the $K^-$
mesons with the same momentum  is equal to be 800 at ion beam
energy 2 AGeV which is far enough below the free $NN$ threshold
($T_{NN}=2.50$ GeV). Due to the Lorentz boost all kaons and
antikaons produced in full solid angle with momenta $< 300$ MeV/c
relative to the projectile nucleus rest frame will be concentrated
in the laboratory inside rather narrow cone $10-20^o$ what
corresponds to the solid angle of 1-3\% of $4\pi$. Note that  the
multiple scattering effect on the detected particles is
significantly decreased in the $Ap$ kinematics due to upward shift 
of both kaon momentum and velocity. Thus, the inverse kinematics is 
well suited for the experimental study of the mechanisms of deep 
subthreshold low-momentum strange meson production.

The feasibility of the experiments discussed above depends on
backgrounds. Subthreshold $K^+$  and $K^-$ meson production is
accompanied by the background of secondary pions and protons which
is much more intense.  Note that the modern magnetic spectrometers
provide reliable $K/\pi$ and $K/p$ separation up to values of
these ratios $10^{-6} - 10^{-7}$ \cite{Ref15,Ref22}. Another
source of background is a particle production from an envelops of
hydrogen target. Usually "target full-target empty" measurements
have to be carried out to obtain the cross sections on hydrogen.
However, this background can be totally removed by using the
windowless target consisting of frozen hydrogen pellets
\cite{Ref24}.

It is worth notice that the method discussed opens the new way to
explore the properties of low-momentum $\eta, \eta',\omega, \rho$
and $\phi$ mesons in nuclear matter, the topic extensively
discussed over the last years \cite{Ref25}. The
experiments in the inverse kinematics may be carried out at
GSI-SIS using an ion beams in 1-2 AGeV region and at ITEP where an
ion beams of energy up to 4.3 AGeV are expected to be available in
the near future.


\end{document}